# Framework for Opinion Mining Approach to Augment Education System Performance

**Amritpal Kaur**
Department of Computer Science and Engineering, Thapar Institute of Engineering and Technology,
(Deemed to be University, Patiala),India
E-mail: akaur5_be15@thapar.edu
**Harkiran Kaur**
Department of Computer Science and Engineering, Thapar Institute of Engineering and Technology,
(Deemed to be University, Patiala),India
E-mail: harkiran.kaur@thapar.edu

**Abstract-** The extensive expansion growth of social networking sites allows the people to share their views and experiences freely with their peers on internet. Due to this, huge amount of data is generated on everyday basis which can be used for the opinion mining to extract the views of people in a particular field. Opinion mining finds its applications in many areas such as Tourism, Politics, education and entertainment, etc. It has not been extensively implemented in area of education system. This paper discusses the malpractices in the present examination system. In the present scenario, Opinion mining is vastly used for decision making. The authors of this paper have designed a framework by applying Naïve Bayes approach to the education dataset. The various phases of Naïve Bayes approach include three steps: conversion of data into frequency table, making classes of dataset and apply the Naïve Bayes algorithm equation to calculate the probabilities of classes. Finally the highest probability class is the outcome of this prediction. These predictions are used to make improvements in the education system and help to provide better education.

## 1. INTRODUCTION

Micro blogging and social media sites have become most easy and common means of communication among peer users. In the recent years, it has been observed that social networking sites such Facebook, Twitter and other similar sites have made a huge impact on people's life and activities. It is observed that people feel free to share their opinions on the internet. Billions of web clients are utilizing social sites to extract the opinions of people on various topics. The process of extracting the opinions of the people, analyze then and classify the sentiment behind these opinions is called Opinion Mining.

Opinion Mining (OM) is the area of Natural Language Processing (NLP) and Artificial Intelligence (AI) for extracting the views about particular topic and classify it as positive, negative or neutral based on people's emotions and sentiments [4]. This is used for decision making in various fields such as finding client satisfaction, reputation of the organization in the market, enhancing the quality of the particular product manufactured by these organization.

Opinion mining modeling is done by three approaches –

A. *Machine Learning approach(ML):-* Machine Learning approach (ML) is beneficial for classification of opinions whether it represents positive or a negative sentiment. This approach is categorized as: Supervised Machine Learning(SML) SML includes labeled dataset where opinions are labeled with appropriate names. Unsupervised Machine Learning (UML) UML includes unlabeled dataset in which opinions are not properly labeled with appropriate names.

B. *Lexicon Based approach*: - Lexicon Based (LB) approach is based on finding the sentiment lexicon which is utilized for classifying the text. In this either dictionary or corpus based approach is applied by semantic methods.

C. *Hybrid approach:-* Hybrid approach (HB) is the combination of Machine Learning and Lexicon Based approach. It has been proved that this combination gives better performance because it uses the Deep Learning and feature extraction method to increase the efficiency of model.

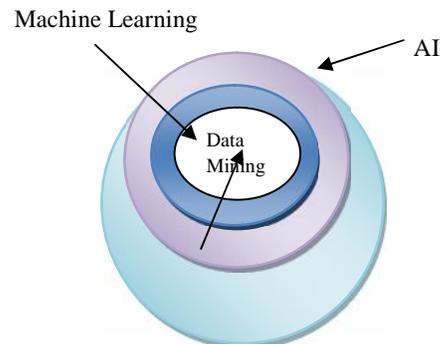

**Fig 1: Layered approach of Data Mining**

In many developing countries such as Fiji and Korea, Government uses sentiment analysis to improve the present education system. Government has taken many decisions in the education field for various parameters such as extracting the opinions of students regarding higher studies, what type of environment they want in educational institutes, how teacher's behavior and method of teaching affects the student's curriculum and performance in academics with the help of this method. Due to this the Government uses Opinion mining as a key point in decision making for educational issues. This paper implements and represents the analysis of the opinion of the students about the examination malpractices among Secondary school students in India. These malpractices have a great impact on the student's performance scale factor. In the proposed students can share their opinion freely on social media regarding this problem, discussing what are reasons to behind these activities, how these affect student's performance.

## 2. RELATED WORK

Gamal D., Et.al in[1] has proposed a technique which demonstrates the research work done on dataset of the social media and illustrates how to improve the efficiency of the models and get more accurate results. Firstly it explained the three approaches for sentiment classification: ML, LB and HB. In this paper various ML algorithms were applied with the different Feature extraction methods to improve efficiency of classification model. It is stated by the authors that Unigram with SVM is more accurate than other algorithms of ML. Feature extraction is important aspect in Sentiment Analysis. Many researchers integrate multiple methods with different features to see the effect on efficiency of the proposed model.

Lavanya K. and Deisy C. [11] gave the idea to extract the data belonging to different domains. It is difficult to train a classifier model so that it can classify the tweets belonging to different domains. Twitter serves as platform for public to share their views about different kinds of products, politics, etc. One classifier performs excellent for one domain but poor for the other domain. Some researchers suggest the SVM to train the data set on feature sets. And SVM gives more accuracy for multi-class data. Data parsing is used to represent words in the form of negative and positive scores. The authors proposed A topic adaptive classifier which can be used to overcome this problem .This classifier uses two vectors- text feature and non text features. This features values are calculated as point wise mutual information and information retrieval. The non features are classified under temporal features, emoticon features and punctuation features.

Napitu F. Et.al [10] stated that, it is very important for companies to maintain their customers in the competitive market. Researchers evaluated a term Churn Management which is used to maintain their valuable customer and predicting customer's churn including complement data and customer usage. Two techniques were followed by the authors in their paper, to predict the model for churn rate as Opinion Finder and Google Profile. Opinion Finder represents customer's opinions as positive and negative words. The technique of Google profile for mood analysis states this technique has 6 dimensions. These include Sure, Kind, vital and Happy., calm and alert. Predictive Model of churn rate requires two inputs a. the past 3 month tweets b. the same combined with frequent time series moods. A recurrent neural network is mostly used to predict the changes in the churn rate.

Adinarayana S.and Ilavarasan E.in [2] proposed an Algorithm – "Over Sampled Imbalance Data Learning(OSIDL)" to retrive the information from the imbalanced data sets and are compared with the data available on twitter corpus by using traditional C4.5 Algorithm. Imbalance dataset exist in the classes which are not balanced. This method is applied by dividing the dataset into majority and minority data-subsets. minority sub set is further processed as: i. First step of OSIDL was constructing improved minority data sets. Then, minority datasets are analyzed to remove the noisy instances to generate pure data set and resample it. After this, the newly minority data set and strong majority dataset were combined to form a single balanced dataset, which increases the accuracy of the algorithm. Also some experiments were conducted by authors to determine the effectiveness of the OSIDL Algorithm, by calculating the precision, recall, TP rate etc.

Zhang Z., Li H. and Yu W. in [3] proposed a paper which explores the two main aspects, i.e. clustering and classification. Clustering analyzes the reviews of customer related to particular attributes such as price aspect and service attitudes. The authors proposed VC-word2vec Algorithm for the clustering. However the sentiment classification divided the customer's sentiment as positive, negative and neutral sentiments. This algorithm used high dimensional words to overall semantics sentences, and then voting algorithm for clustering according to the features words. For conducting the sentiment classification the authors proposed an algorithm based on ED-TextRank, which focus on the selection of aspect features and combines sentiment dictionary with TextRank. This analysis is done for predicting the market value and future demand of the product After applying Various experiments are conducted with these algorithms , ED-TextRank give more accuracy than others.

Zhang T. in [5] studied some case studies and concluded that people used to watch TV program online and gave feedback for different TV programs. Subjective evaluations of TV programs are executed on the basis of satisfaction survey,that analyzes how much user is satisfied by the channels or programs. Objective and subjective evaluations are differentiated by 2-POS Subjective model which is dependent on dictionary ("Corpus"). Then it is evaluated by POS tagging, followed by judging sentiment words orientation. This was done by applying SO-PMI (Sentiment Orientated Point-wise mutual information). It is UML method that measures the relevance of words by computing occurrences of various words. Different experiments are performed to meet the criteria of above algorithms.

Ejaz A., et.al in [6] compared LB approach with n-grams to three models of ML, that is, Random Forest (RF) with word vec Decision Tree, and Random Forest with n-gram. These models are applied on Amazon's product feedback dataset. A sentence having a fact is called objective sentence and sentence having an opinion is called subjective sentence. First of all, data set taken from Amazon's website is pre-processed. In the pre-processing stop-word, punctuation are removed and stemming is done. Then, a model classifier was applied on dataset.. Various experiments are done on different data set by applying these algorithms.

Prameswari P., Surjandari I.,Laoh E. in [7] focus on Bali island which is a popular tourist place. It was needed to make continuous improvement by giving attention to various services. And this could be done by identifying the opinion of the tourists based on online reviews on various websites. To perform above tasks Recursive Neutral Tensor Network (RNTN) was used. As the outcome of this algorithm, various services were improved to attract the more tourists towards Bali Island. The dataset was taken from a trusted website. Then data cleaning and preprocessing was done, followed by applying this algorithm. Based on various experiments the results shown by RNTN were better than the other algorithms.

Soni D., Sharma M., Khatri S. in [8] studied the opinions of people regarding political and non-political issues.it was observed that these issues are directly affected by views shared on social media. In this paper, a simple data set was taken from social sites during election time and opinions of user were analyzed. First of all data cleaning and preprocessed was done and then Logistic regression model was applied on data set and various parameters such as recall, precision rate were calculated to determine the accuracy of the above model.

## 3. COMPARATIVE ANALYSIS OF OPINION MINING APPROACHES

**Table I: Approaches of Opinion mining analysis**

| S.No | Approach | Application area | Outcomes |
|---|---|---|---|
| 1. | SVM model classifier[1] | Social Media | Accuracy of different ML models can be increased by using feature selection |
| 2. | OSIDL[2] | Social Media(Twitter) | Developed a new algorithm OSIDL to get the knowledge from imbalance datasets |
| 3. | VC-Word2vec ED-TextRank[3] | Product Industry | Proposed algorithms performs more accurate experiments than previous available methods |
| 4. | SO-PMI[5] | Social Media | Three steps were proposed: subjective and objective evaluations, to get feature words and opinions |
| 5. | RNTN[7] | Tourism | Classified the sentiment of people about a particular Bali island by using RNTN algorithm. |
| 6. | Customer Churn Rate[10] | Broadband Internet | Proposed a technique Churn management, by using this technique companies can study customer's sentiments. |
| 7. | NaïveBayes[9] | Education | Analysis the student's opinion for education system by using Facebook data and results that NaïveBayes is better algorithms than other algorithms for educational data. |

But the idea to check upon examination malpractices is not discussed by government that how it affects the student's life and what are the reasons for it. This paper includes the analysis of student's opinion that what are the reasons for these malpractices and who is responsible for the same.

## 4. PROPOSED WORK

The problem is malpractices present in examination centre of India. The malpractices include, cheating allowed, leakage of question paper before exams, Un-fair ways of checking answer sheets. This affects the result of hard working students. The analysis of reviews of students is discussed in this paper to overcome this problem.

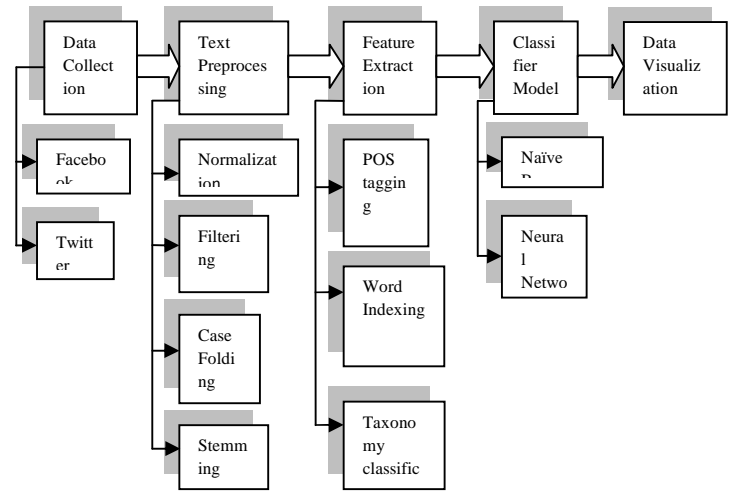

**Fig 2. Different phases of implementation**

The proposed system of Opinion mining will be implemented by using the combination of tools Python and Natural Language Processing (NLP). Python has some in-built libraries but Natural Language Toolkit (NLTK) provides a platform to create python programs to interact with natural data and take decisions accordingly.

In this section, a framework is provided which describes that how data is extracted from social sites, then preprocess the data and classify the opinions as positive, negative and neutral. [case study]This process has 5 following steps:

*A. Feedback collection*: Students and Teacher's review is taken by posting a questionnaire on social sites such as Facebook, Twitter. This questionnaire contains the basic questions such as:

a) what are reasons for the malpractices in examination control board

b) what are the effects of the same on student's dedication towards the hard work. Twitter data was extracted using the Twitter Streaming API(web scraping). Keywords related to education such as curriculum reforms, education were used for keywords for streaming and retrieving the feedback from website. The feedback extracted from website can be kept in JSON format (JavaScript Object Notation) so that preprocessing steps can be implement on the educational dataset.

*B. Text Pre-processing*: After extracting the data from the social sites it requires a series of pre-processing steps applied before the data is used for Text Mining. Text pre-processing contains the several steps which are applied according to needs of the case study. These steps which are used in this study of this dataset are normalization spellings, remove missing values, case folding, filtering, lemmatization and stemming. Table II describes these steps of text preprocessing:

**Table II : Steps of Text Pre-Processing**

| S.No. | Task | Description |
|---|---|---|
| 1. | Spell checker | The process of correcting misspelled words. |
| 2. | Stopword Removal | The process of removing meaningless words, punctuations and 'to', 'the' are also removed by step |
| 3. | Case Folding | make the whole document in one form. |
| 4. | Stemming | The process of reducing the derived words into their stem (root word) e.g. the combination word has the root word combine. |

*C. Aspect Extraction* : Aspect Extraction is simply stated as reprocessing the data and take out the essential features from the data. This can be done by various steps such as removing all the words which do not begin with alphabetic order, it simply means that removing the words starting with Numbers or with Special characters. Next step is removing stop words such as is, am, are present in the text. After it various steps such as POS tagging, Word Indexing, Taxonomy Formulation are performed to extract feature. Table III describes the various stages:

**Table III : Stages of Feature Extraction**

| No. | Stage | Description |
|---|---|---|
| 1 | POS tagging | It is the process to find nouns and determining the semantic of words present in text. |
| 2 | Noun Indexing | The process of indexing nouns in the text. |
| 3 | Taxonomy Formulation | It is the process of grouping the nouns into categories that are present in the dataset. |

*D. Classifier Model:-* This step includes two stages: Training and Classification. In training stage, the model is trained by using Naïve Bayes algorithm on some part of the dataset. In classification that trained model is used to predict the outcomes. Naive Bayes is used because it is observed in various experiments this algorithm gives more accurate results with the educational dataset. As this model gives more accuracy that's why it is normally used . Equation 1 shows the Naïve Bayes for opinion mining[12]:

$$P(s|M) = \frac{P(s).P(M|s)}{P(M)} \quad (1)$$

Where s is a opinion. M is a comment. P(s) is the probability of a opinion. P(M|s) is the probability that a given comment is being classified as a opinion . P(M) is the probability that it is actually a comment.

*E. Analysis the results:-*The basic purpose of the analysis is to get useful information from the feedback for better interpretation and understanding student's emotions regarding educational system especially about malpractices in the Examination Control System. Government can use this analysis to give idea to check upon examination system so that they can take better decision for the students, and quality of education can be improved. To attain this purpose, various visualization tools such as charts, graphs can applied on the results of opinion mining. It will help the Government in decision making and save time of the communities.

## 7. CONCLUSION

Opinion Mining has great potential in the field of education. It provides an analysis which is very helpful in decision making. This paper showcases Student's feedback on malpractices in the examination system (source: Facebook, Twitter). In this paper, NaïveBayes Algorithm is suggested to analyze the opinions on malpractices in education system because this algorithm is well suited for the education dataset. The proposed technique is used to overcome with the un-fair means of conducting examinations. This will help the Government to get more ideas by analyzing the outcome of this technique and implement new rules and regulations for minimizing the malpractices in examination control system.